\documentclass[twocolumn,floatfix, showpacs]{revtex4}

\usepackage[final]{epsfig}
\usepackage{amsmath}
\usepackage{parskip}
 
\begin{document}
 
\title{The Nuclear Suppression Factor for $P_T$ increasing towards the Kinematic Limit}
 
\author{Thorsten Renk}
\email{thorsten.i.renk@jyu.fi}
\affiliation{Department of Physics, P.O. Box 35 FI-40014 University of Jyv\"askyl\"a, Finland}
\affiliation{Helsinki Institute of Physics, P.O. Box 64 FI-00014, University of Helsinki, Finland}
 

\begin{abstract}
The suppression of high transverse momentum ($P_T$) hadron production in ultrarelativistic heavy-ion (A-A) collisions as compared to the scaled expectation from proton-proton (p-p) collisions expressed as the nuclear modification factor $R_{AA}$  is experimentally well established and can be traced back to interactions between the hard parton shower and the soft bulk matter. Physical intuition suggests that the medium modification should cease to be important when the hard scale of parton production is much larger than the typical momentum scale in the medium (e.g. the temperature $T$) and that consequently $R_{AA}(P_T) \rightarrow 1$ for $P_T \gg T$. However, $R_{AA}$ is not a 'simple' observable, but rather results from many different mechanisms which influence what happens when $P_T$ increases.  In particular, $R_{AA}$ does not necessarily approch unity even if the hadron momentum is probed at the kinematic limit of the reaction. The aim of this work is to identify and discuss such mechanisms and to present different model expectations of what one would find if one could measure suppression out to the kinematic limit for hard hadron production and give predictions for the $P_T$ dependence at both RHIC and LHC.
\end{abstract}
 
\maketitle

\section{Introduction}

Jet quenching, i.e.\ the energy loss of hard partons created in the first moments of an A-A collision due to interactions with the surrounding soft medium  has long been regarded a promising tool to study properties of the soft bulk medium created along with the hard process \cite{Jet1,Jet2,Jet3,Jet4,Jet5,Jet6}. The effect of the medium is apparent from a comparison of high $P_T$ hadron observables  measured in A-A collisions with the same observables in p-p collisions. The current range of such observables includes the suppression in single inclusive hard hadron spectra $R_{AA}$ \cite{PHENIX_R_AA}, the suppression of back-to-back correlations \cite{Dijets1,Dijets2} and single hadron suppression as a function of the emission angle with the reaction plane \cite{PHENIX-RP}. Recently also preliminary measurements of fully reconstructed jets have become available \cite{STARJET}.

Single hadron observables and back-to-back correlations above 6 GeV (where hadron production is dominated by hard processes) are well described in detailed model calculations using the concept of energy loss \cite{HydroJet1,Dihadron1,Dihadron2}, i.e. under the assumption that the process can be described by a medium-induced shift of the leading parton energy by an amount $\Delta E$ where the probability of energy loss is governed by a distribution $P(\Delta E)$, followed by a fragmentation process using vacuum fragmentation of a parton with the reduced energy. This can be cast into the form of a modified fragmentation function (MMFF). If the vacuum fragmentation function, i.e. the distribution of hadrons produced from a parton at fractional momentum $z$ given a hadronization scale $\mu^2$ is $D(z,\mu^2)$, then the MMFF given the medium induced energy loss probability $P(\Delta E)$ can be written as

\begin{equation}
\label{E-ModF}
\tilde{D}(z,\mu^2) = \int_0^E d \Delta E  P(\Delta E) \frac{D\left( \frac{z}{1-\Delta E/E},\mu^2\right)}{1-\Delta E/E}.
\end{equation}

Beyond the leading parton approximation in which energy loss and fragmentation factorize, one has to solve the full partonic shower evolution equations in the medium while assuming that the non-perturbative hadronization takes place outside the medium. At least for light subleading hadrons in a shower, factorizing hadronization from the medium-modified parton shower is a reasonable assumption at both RHIC and LHC kinematics. There are several calculations which utilize such medium-modified showers analytically \cite{HydroJet2,HydroJet3,Dihadron3}. Recently, also Monte Carlo (MC) codes for in-medium shower evolution have become available \cite{JEWEL,YAS,YAS2,Carlos,Carlos2,Martini} which are based on MC shower simulations developed for hadronic collisions, such as PYTHIA \cite{PYTHIA} or HERWIG \cite{HERWIG}. These have, unlike current analytical computations, full energy-momentum conservation enforced at each branching vertex. In these calculations the MMFF is obtained directly rather than from an expression like Eq.~(\ref{E-ModF}).

So far, the different pictures for the parton-medium interaction have been explored and compared with data over a rather finite kinematical window with $P_T < 20$ GeV. There is a widespread expectation that if the $P_T$ range of the measurement could be extended, either at RHIC or at LHC, one would eventually observe the disappearance of the medium effect. The origin of this expectation is that the medium is able to modify the hard parton kinematics at a typical scale set by its temperature $T$, whereas the parton dynamics takes place at a partonic hard scale $p_T$, and if $p_T \gg T$ the hard kinematics should be esentially unchanged, which can be realized for large hadronic $P_T$. For example, in the case of the nuclear suppression factor, this expectation would imply that $R_{AA}(P_T)$ approaches unity for $P_T \gg T$. It is the aim of this paper to discuss the physics contained in the shape of $R_{AA}(P_T)$ and to present what current models, both based on the energy-loss concept and on the in-medium parton shower concept, predict for the shape of $R_{AA}$ at very large momenta at both RHIC and LHC.

\section{Nuclear suppression in the energy loss picture}

\subsection{Qualitative estimates}

A qualitative argument why $R_{AA}$ should increase with $P_T$ can be made as follows: Parton spectra can be approximated by a power law as $dN/dp_T = const./p_T^n$ where $n\approx 7$. Assume that one can approximate the effect of the medium by the mean value energy loss $\langle \Delta E \rangle$ (for realistic energy loss models, this is not a good approximation, as fluctuations around the mean turn out to be large). In this case, the energy loss shifts the spectrum. This can be described by the replacement $p_T \rightarrow p_T + \langle \Delta E \rangle$ in the expression for the parton spectrum. $R_{AA}(p_T)$ can then be approximated by the ratio of the parton spectra before and after energy loss as

\begin{equation}
\label{E-RAAApprox1}
R_{AA}(p_T) \approx \left(\frac{p_T}{p_T + \langle \Delta E \rangle}\right)^n = \left(1 - \frac{\langle \Delta E \rangle}{p_T + \langle \Delta E \rangle}\right)^n
\end{equation}

and it is easily seen that this expression approaches unity for $p_T \gg \langle \Delta E\rangle$. However, it is not readily obvious under what conditions the limit is reached even if the medium properties are known. Parametrically, the medium temperature $T$ governs both the medium density and the typical medium momentum scale, but the total energy loss represents the cumulative effect of the medium, i.e. a line integral of medium properties along the path of the partons, and furthermore the physics of medium-induced radiation is rather complicated, interference between different radiation graphs play a significant role, and therefore the mean energy loss is not simply $\sim  T$. Thus, in realistic calculation the mean energy loss at RHIC conditions is $\langle \Delta E \rangle \approx O(10)$ GeV even for $T < 0.35$ GeV \cite{Dihadron2}, and hence it can be understood that current data are relatively far from the limit.

There are five main points which may be raised against the approximation Eq.~(\ref{E-RAAApprox1}):

\begin{itemize}
\item the estimate holds for partons and does not take into account fragmentation:

This, however, is not a crucial issue for the question at hand. The fragmentation function $D(z,\mu^2)$ is steeply falling with $z$ and as a result, fragmentation processes at low $z$ are preferred. However, for given momentum scale of the hadron spectrum, low $z$ implies high parton momentum, and this is suppressed because the parton spectrum is also steeply falling with $p_T$. As a result, there is some typical intermediate $\langle z \rangle$ (dependent on hadron and parton type) which relates hadron and parton momentum, for quarks fragmenting into light hadrons at RHIC kinematics $\langle z \rangle \approx 0.5-0.7$. This means that hadronic $R_{AA}$ is to first approximation simply scaled by this factor as compared to partonic $R_{AA}$. Fluctuations around the average tend to smear out structures in the partonic $R_{AA}$ through the hadronization process, but do not alter the shape of $R_{AA}(P_T)$ beyond that. Thus, qualitatively Eq.~(\ref{E-RAAApprox1}) holds also on the hadronic level.

\item the estimate does not distinguish between quarks and gluons:

This is moderately important, as energy loss is expected to be stronger for gluons by a factor 9/4 (the ratio of the Casimir color factors). At low $P_T$, hadron production is driven by gluonic processes as gluons are copiously available in the low $x$ region in parton distribution functions (PDFs) \cite{CTEQ1,CTEQ2,NPDF,EKS98,EPS09}. However, hadron production at higher $P_T$ probes more and more in the high $x$ region in the parton distributions, and eventually valence quark scattering dominates. The hadronic $R_{AA}$ should therefore show a rise from gluonic $R_{AA}$ to the larger value of quark $R_{AA}$ which corresponds to the transition from gluon to quark-dominated hadroproduction. As shown in \cite{RAA_Proton}, this is likely to be the mechanism underlying the rising trend observed in $R_{AA}$ at RHIC. For asymptotically high energies, the mechanism is not relevant however, as this is always a quark dominated regime.

\item the estimate neglects fluctuations around the average energy loss:

In the presence of fluctuations, $P(\Delta E)$ can be written as the sum of three terms, corresponding to transmission witout energy loss, shift of the parton energy by finite energy loss or parton absorption as

\begin{equation}
P(\Delta E) = \tilde{T} \delta(\Delta E) + \tilde{S} \cdot \tilde{P}(\Delta E) + \tilde{A} \cdot \delta(\Delta E - E)
\end{equation}

where $\tilde{P}(\Delta E)$ is a normalized probability distribution such that $\tilde{T}+\tilde{S}+\tilde{A}=1$. Inserting this form to average over Eq.~(\ref{E-RAAApprox1}) with the proper weights, one finds

\begin{equation}
R_{AA} \approx \tilde{T} +  \int d\Delta E \cdot \tilde{S} \cdot \tilde{P}(\Delta E)  \left(1-\frac{\Delta E}{p_T + \Delta E}\right)^n.
\end{equation}

It follows that $R_{AA}$ obtained with this expression is always bounded by $\tilde{T}$ from below (if a fraction of partons escapes unmodified, no amount of modification to the rest will alter this) and by $(1-\tilde{A})$ from above (if partons are absorbed {\em independent of their energy}, $R_{AA}$ will never approach unity). In many calculations, $\tilde{A}$ is determined by the condition that a parton is absorbed whenever its calculated energy loss exceeds its energy, i.e. $\tilde{A}$ and $\tilde{S}$ are dependent on the initial parton energy. In particular, in the ASW formalism \cite{QuenchingWeights}, the energy loss can be formally larger than the initial energy, since the formalism is derived for asymptotically high energies $E\rightarrow \infty$ and small energy of radiated gluons $\Delta E \ll E$, but is commonly applied to kinematic situations in which these conditions are not fulfilled.
$R_{AA}$ at given $p_T$ is then equal to the transmission term $\tilde{T}$ plus a contribution which is proportional to the integral of $\tilde{P}(\Delta E)$ from zero up to the energy scale $E_{max}$ of the parton, {\it seen through the filter} of the steeply falling parton spectrum. Thus, in the presence of fluctuations, $R_{AA}$ is dominated by fluctuations towards the low $\Delta E$, and  $R_{AA}$ grows with $p_T$ since $E_{max}$ grows linearly with $p_T$. If $P(\Delta E)$ includes fluctuations up to a maximal energy loss $\Delta E_{max}$, then for $p_T \gg \Delta E_{max}$ the original argument made for constant energy loss applies and $R_{AA}$ aproaches unity. In practice this may not be observable - the energy loss probability for RHIC kinematics may be substantial up to scales of $O(100)$ GeV \cite{Dihadron2}, i.e. of the order of the kinematic limit.

\item the pQCD parton spectrum is not a power law:

\begin{figure}
\epsfig{file=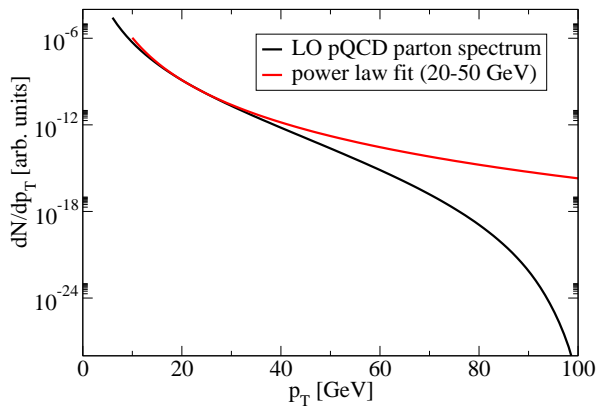, width=7.8cm}
\caption{\label{F-Power}(Color online) The LO pQCD parton spectrum for $\sqrt{s}=200$ GeV compared with a power law fit to the region from 20 to 50 GeV.}
\end{figure} 

While in a limited kinematic range, the pQCD parton spectrum is approximated well by a power law, at about $\sqrt{s}/4$ the power law fit becomes a bad description of the spectrum. This is shown in Fig.~\ref{F-Power}. Close to the kinematic limit at $\sqrt{s}/2$, the parton spectrum falls very steeply. If one would attempt a local power law fit in this region, the region of validity for the fit would be small and $n$ very large. One can readily see from Eq.~(\ref{E-RAAApprox1}) that even for $p_T \gg \langle \Delta E \rangle$ $R_{AA}$ does not approach unity when at the same time $n \rightarrow \infty$. In other words, close to the kinematic limit, even a small $\Delta E$ causes a massive suppression simply because there are no partons available higher up in the spectrum which could be shifted down. For this reason, close to the kinematic limit $R_{AA} \rightarrow 1$ can not be expected, rather (dependent on the details of modelling), something like $R_{AA} \rightarrow \tilde{T}$ should be expected.

However, note also that the validity of factorization into a hard process and a fragmentation function has been assumed for hadron production up to the kinematic limit. This may not be true, Higher Twist mechanisms like direct hadron production in the hard subprocess (in the context of heavy-ion collisions, see e.g. \cite{DirectHP} may represent a different contribution which, due to color transparency, remains unaffected by the medium at all $P_T$ and which may be significantly stronger than fragmentation close to the kinematic limit. This could be effectively absorbed into a modified coefficient $\tilde{T}$ which however ceases to have a probabilistic interpretation.

\item the nuclear initial state effects have not been taken into account:

The initial state nuclear effects, i.e. the difference in nucleon \cite{CTEQ1,CTEQ2} and nuclear \cite{NPDF,EKS98,EPS09} PDFs, are often thought to be a small correction to the final state medium effects. Over a large kinematic range, that is quite true. However, as one approaches the kinematic limit and forces the distributions into the $x \rightarrow 1$ valence quark distributions, one probes the Fermi motion region in the nuclear parton distributions where the difference to nucleon PDFs is sizeable.

\begin{figure}
\epsfig{file=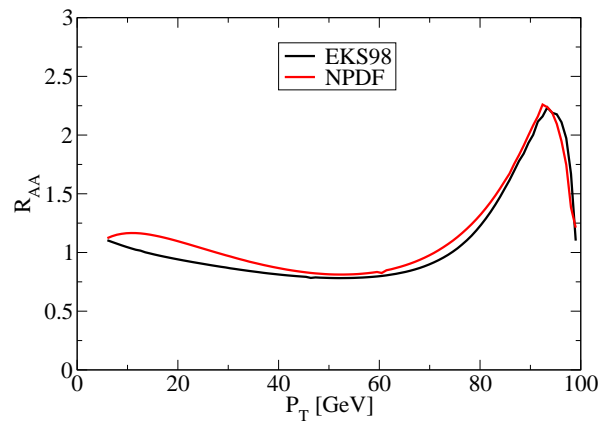, width=7.8cm}
\caption{\label{F-NPDF}(Color online) $R_{AA}(P_T)$ calculated with nuclear {\em initial state} effects only, obtained from the NPDF set \cite{NPDF} and the EKS98 \cite{EKS98} set of nuclear parton distributions calculated for the whole kinematic range at RHIC.}
\end{figure}

In Fig.~\ref{F-NPDF}, $R_{AA}(P_T)$ is shown for RHIC kinematics taking into account only the nuclear initial state effects with two different sets of nuclear PDFs, but no final state medium induced energy loss. It is readily apparent that over most of the kinematic range, $R_{AA}(P_T) \approx 1$, but that there is a strong enhancement visible above 80 GeV.

\end{itemize}

\subsection{Detailed calculation}

The detailed calculation of $R_{AA}$ in the energy loss models presented here follows the Baier-Dokshitzer-Mueller-Peigne-Schiff (BDMPS) formalism for radiative energy loss  \cite{Jet2} using quenching weights as introduced by Salgado and Wiedemann \cite{QuenchingWeights}, commonly referred to as the Armesto-Salgado-Wiedemann (ASW) formalism. 

The probability density $P(x_0, y_0)$ for finding a hard vertex at the transverse position ${\bf r_0} = (x_0,y_0)$ and impact parameter ${\bf b}$ is given by the product of the nuclear profile functions as

\begin{equation}
\label{E-Profile}
P(x_0,y_0) = \frac{T_{A}({\bf r_0 + b/2}) T_A(\bf r_0 - b/2)}{T_{AA}({\bf b})},
\end{equation}

where the thickness function is given in terms of Woods-Saxon the nuclear density
$\rho_{A}({\bf r},z)$ as $T_{A}({\bf r})=\int dz \rho_{A}({\bf r},z)$ and $T_{AA}({\bf b})$ is the standard nuclear overlap function $T_{AA}({\bf b}) = d^2 {\bf s}\, T_A({\bf s}) T_A({\bf s}-{\bf b})$. 

If the angle between outgoing parton and the reaction plane is $\phi$, the path of a given parton through the medium $\zeta(\tau)$, i.e. its trajectory $\zeta$ as a function of proper medium evolution time $\tau$ is determined in an eikonal approximation by its initial position ${\bf r_0}$  and the angle $\phi$ as $\zeta(\tau) = \left(x_0 + \tau \cos(\phi), y_0 + \tau \sin(\phi)\right)$ where the parton is assumed to move with the speed of light $c=1$ and the $x$-direction is chosen to be in the reaction plane. The energy loss probability $P(\Delta E)_{path}$ for this path can be obtained by evaluating the line integrals along the eikonal parton path

\begin{equation}
\label{E-omega}
\omega_c({\bf r_0}, \phi) = \int_0^\infty \negthickspace d \zeta \zeta \hat{q}(\zeta) \quad  \text{and} \quad \langle\hat{q}L\rangle ({\bf r_0}, \phi) = \int_0^\infty \negthickspace d \zeta \hat{q}(\zeta)
\end{equation}

with the relation 

\begin{equation}
\label{E-qhat}
\hat{q}(\zeta) = K \cdot 2 \cdot \epsilon^{3/4}(\zeta) (\cosh \rho - \sinh \rho \cos\alpha)
\end{equation}
assumed between the local transport coefficient $\hat{q}(\zeta)$ (specifying the quenching power of the medium), the energy density $\epsilon$ and the local flow rapidity $\rho$ with angle $\alpha$ between flow and parton trajectory \cite{Flow1,Flow2}. The medium parameters $\epsilon$ and $\rho$ are obtained from a 2+1-d hydrodynamical simulation of bulk matter evolution \cite{Hydro}, chosen to have the RHIC and the LHC medium described within the same framework. $\omega_c$ is the characteristic gluon frequency, setting the scale of the energy loss probability distribution, and $\langle \hat{q} L\rangle$ is a measure of the path-length weighted by the local quenching power. The parameter $K$ is seen as a tool to account for the uncertainty in the selection of the strong coupling $\alpha_s$ and possible non-perturbative effects increasing the quenching power of the medium (see discussion in \cite{Dijets2}) and adjusted such that pionic $R_{AA}$ for central Au-Au collisions is described at one value of $P_T$.

Using the numerical results of \cite{QuenchingWeights} and the definitions above, the energy loss probability distribution given a parton trajectory can now be obtained as a function of the initial vertex and direction $({\bf r_0},\phi)$ as $P(\Delta E; \omega_c({\bf r},\phi), R({\bf r},\phi))_{path} \equiv P(\Delta E)_{path}$ for $\omega_c$ and $R=2\omega_c^2/\langle\hat{q}L\rangle$. From the energy loss distribution given a single path, one can define the averaged energy loss probability distribution $P(\Delta E)\rangle_{T_{AA}}$ as

\begin{equation}
\label{E-P_TAA}
\langle P(\Delta E)\rangle_{T_{AA}} \negthickspace = \negthickspace \frac{1}{2\pi} \int_0^{2\pi}  
\negthickspace \negthickspace \negthickspace d\phi 
\int_{-\infty}^{\infty} \negthickspace \negthickspace \negthickspace \negthickspace dx_0 
\int_{-\infty}^{\infty} \negthickspace \negthickspace \negthickspace \negthickspace dy_0 P(x_0,y_0)  
P(\Delta E)_{path}.
\end{equation}

The energy loss probability  $P(\Delta E)_{path}$ is derived in the limit of infinite parton energy  \cite{QuenchingWeights}, however in the following the formalism is applied to finite kinematics. In order to account for the finite energy $E$ of the partons $\langle P(\Delta E) \rangle_{T_{AA}}$ is truncated
at $\Delta E = E$ and $\delta(\Delta E-E) \int^\infty_{E} d\Delta E \,P(\Delta E)$ is added to the truncated distribution to ensure proper normalization. The physics meaning of this correction is that all partons are considered absorbed if their energy loss is formally larger than their initial energy. The momentum spectrum of hard partons is calculated in leading order perturbative Quantum Chromodynamics (LO pQCD) (explicit expressions are given in \cite{Dijets2} and references therein). The medium-modified perturbative production of hadrons can then be computed from the expression
\begin{equation}
d\sigma_{med}^{AA\rightarrow h+X} \negthickspace \negthickspace = \sum_f d\sigma_{vac}^{AA \rightarrow f +X} \otimes \langle P(\Delta E)\rangle_{T_{AA}} \otimes
D^{f \rightarrow h}(z, \mu^2)
\end{equation} 
where $d\sigma_{vac}^{AA \rightarrow f +X}$ is the partonic cross section for the inclusive production of a parton $f$, $D^{f \rightarrow h}(z, \mu^2)$ the vacuum fragmentation function for the hadronization of a parton $f$ into a hadron $h$ with momentum fraction $z$ and hadronization scale $\mu$ and from this the nuclear modification factor $R_{AA}$ follows as
\begin{equation}
\label{E-RAA}
R_{AA}(p_T,y) = \frac{dN^h_{AA}/dp_Tdy }{T_{AA}({\bf b}) d\sigma^{pp}/dp_Tdy}.
\end{equation}

\section{Nuclear suppression in the medium-modified shower picture}

\subsection{Qualitative arguments}

In a medium-modified shower picture, the whole partonic in-medium evolution of a parton shower following a hard process is studied, leading to a modification of the fragmentation function (FF) which is more general than Eq.~(\ref{E-ModF}). In this framework, $R_{AA} \rightarrow 1$ is realized if the MMFF becomes sufficiently similar to the vacuum FF. A qualitative argument why the MMFF should approach the vacuum FF for $P_T \gg T$ can be made by considering for example the RAD (radiative energy loss) scenario of the MC code YaJEM (Yet another Jet Energy-loss Model). This model is described in detail in Refs. \cite{YAS,YAS2}. 

The parton shower developing from a highly virtual initial hard parton in this model is described as a series of $1\rightarrow 2$ splittings $a \rightarrow bc$ where the virtuality scale decreases in each splitting, i.e. $Q_a > Q_b,Q_c$ and the energy is shared among the daugther partons $b,c$ as $E_b = z E_a$ and $E_c = (1-z) E_a$. The splitting probabilities for a parton $a$ in terms of $Q_a, E_a$ are calculable in pQCD and the resulting shower is computed event by event in a MC framework.  
In the absence of a medium, the evolution of the shower is obtained using the PYSHOW routine \cite{PYSHOW} which is part of the PYTHIA package \cite{PYTHIA}.

In the presence of a medium, the main assumption of YaJEM is that the parton kinematics or the splitting probability is modified. In the RAD scenario, the relevant modification is a virtuality gain

\begin{equation}
\label{E-Qgain}
\Delta Q_a^2 = \int_{\tau_a^0}^{\tau_a^0 + \tau_a} d\zeta \hat{q}(\zeta)
\end{equation}

of a parton during its lifetime through the interaction with the medium. In order to evaluate Eq.~(\ref{E-Qgain}) during the shower evolution, the momentum space variables of the shower evolution equations need to be linked with a spacetime position in the medium. This is done via the uncertainty relation for the average formation time as

\begin{equation}
\label{E-Lifetime}
\langle \tau_b \rangle = \frac{E_b}{Q_b^2} - \frac{E_b}{Q_a^2}
\end{equation} 

and randomized splitting by splitting by sampling from the distribution

\begin{equation}
\label{E-RLifetime}
P(\tau_b) = \exp\left[- \frac{\tau_b}{\langle \tau_b \rangle}  \right].
\end{equation}

The limit in which the medium modification is unimportant is then given by $Q^2 \gg \Delta Q^2$, i.e. if the influence of the medium on the parton virtuality is small compared with the virtuality itself, the evolution of the shower takes place as in vacuum. Note that there is always a kinematical region in which the condition can never be fulfilled: The region $z \rightarrow 1$ in the fragmentation function represents showers in which there has been essentially no splitting. Since the initial virtuality determines the amount of branchings in the shower, this means one probes events in which the initial virtuality $Q_0$ is not (as in typcial events) of the order of the initial parton energy $E_0$, but rather $Q_0 \sim m_h$ where $m_h$ is a hadron mass. Since $m_h$ is, at least for light hadrons, of the order of the medium temperature, $Q^2 \gg \Delta Q^2$ can not be fulfilled in the region $z \approx 1$ of the MMFF --- here the medium effect is always visible.

\begin{figure}
\epsfig{file=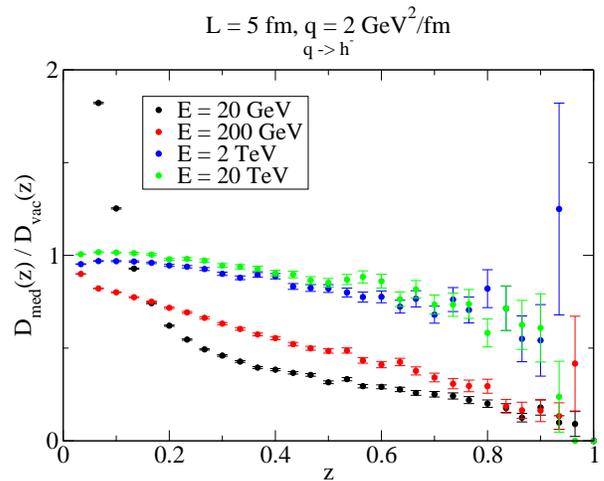, width=7.8cm}
\caption{\label{F-FFnew}(Color online) Ratio of medium-modified over vacuum quark fragmentation function into charged hadrons obtained in YaJEM for a constant medium with 5 fm length and $\hat{q}=2$ GeV$^2$/fm. Shown are the results for different initial quark energies $E$.}
\end{figure} 

This is illustrated in Fig.~\ref{F-FFnew}. Here the ratio of medium-modified over vacuum fragmentation function $D^{q\rightarrow h^-}(z,\mu_p^2)$ as obtained in YaJEM is shown for a constant medium for different initial partonic scales $\mu_p \equiv E$. For a low initial scale of $E=20$ GeV, one observes that the whole range between $z=0.2$ and $z=1$ is suppressed in the medium, whereas the region below $z=0.1$ shows enhancement due to the hadronization of the additional medium-induced radiation. For larger initial scales, the region of enhancement becomes confined to smaller and smaller $z$ and the fragmentation function ratio approaches unity across a large range. However, in the region $z \approx 1$ suppression due to the medium always persists as expected.

As a consequence, one can expect $R_{AA} \rightarrow 1$ for $P_T \gg T$ (where $\Delta Q^2$ is assumed to be parametrically $O(T^2)$) except near the kinematic limit $P_T \approx \sqrt{s}/2$ where the region $z\approx 1$ of the MMFF is probed and suppression is expected to persist.

\subsection{Detailed calculation}

The detailed computation of $R_{AA}$ within YaJEM is outlined in \cite{YAS2}. It shares many steps with the computation within the energy loss picture as described above, in particular the medium averaging procedure.

The basic quantity to compute is the MMFF, given a path through the medium. Due to an approximate scaling law identified in \cite{YAS}, it is sufficient to compute the line integral

\begin{equation}
\label{E-Qsq}
\Delta Q^2_{tot} = \int d \zeta \hat{q}(\zeta)
\end{equation}

in the medium to obtain the full MMFF $D_{MM}(z, \mu_p^2,\zeta)$ from a YaJEM simulation for a given eikonal path of the shower-initiating parton, where $\mu_p^2$ is the {\em partonic} scale. The link between $\hat{q}$ and medium parameters is given as previously by Eq.~(\ref{E-qhat}), albeit with a different numerical value for $K$. The medium-averaged MMFF is then computed as

\begin{equation}
\label{E-D_TAA}
\begin{split}
\langle D_{MM}&(z,\mu_p^2)\rangle_{T_{AA}} \negthickspace =\\ &\negthickspace \frac{1}{2\pi} \int_0^{2\pi}  
\negthickspace \negthickspace \negthickspace d\phi 
\int_{-\infty}^{\infty} \negthickspace \negthickspace \negthickspace \negthickspace dx_0 
\int_{-\infty}^{\infty} \negthickspace \negthickspace \negthickspace \negthickspace dy_0 P(x_0,y_0)  
D_{MM}(z, \mu_p^2,\zeta).
\end{split}
\end{equation}

From this, the medium-modified production of hadrons is obtained from

\begin{equation}
\label{E-Conv}
d\sigma_{med}^{AA\rightarrow h+X} \negthickspace \negthickspace = \sum_f d\sigma_{vac}^{AA \rightarrow f +X} \otimes \langle D_{MM}(z,\mu_p^2)\rangle_{T_{AA}}
\end{equation} 

and finally $R_{AA}$ via Eq.~(\ref{E-RAA}). A crucial issue when computing $R_{AA}$ for a large momentum range is that YaJEM provides the MMFF for a given {\em partonic} scale whereas a factorized QCD expression like Eq.~(\ref{E-Conv}) utilizes a fragmentation function at givem {\em hadronic scale}. In previous publications \cite{YAS,YAS2}, the problem has been commented on, but not addressed, as the variation in momentum scale for current observables is not substantial. In this paper, the matching between partonic and hadronic scale is done as follows:

For several partonic scales, $\langle D_{MM}(z,\mu_p^2)\rangle_{T_{AA}}$ is computed, and the exponent $n$ of a power law fit to the parton spectrum at scale $\mu_p$ is determined. The maximum of $z^n \langle D_{MM}(z,\mu_p^2)\rangle_{T_{AA}}$ corresponds to the most likely value $\tilde{z}$ in the fragmentation process, and thus the partonic scale choice is best for a hadronic scale $P_T = \tilde{z}\mu_p$. The hadronic $R_{AA}$ is then computed by interpolation between different optimal scale choices from runs with different partonic scales. Finally, in the region $P_T \rightarrow \sqrt{s}/2$, $\langle D_{MM}(z,s/4)\rangle_{T_{AA}}$ is always the dominant contribution to hadron production.

The matching procedure between hadronic and partonic scale choice also leads to a significant improvement in the description of $R_{AA}$ in the measured momentum range at RHIC as compared to previous results \cite{YAS,YAS2}.

\section{Results for RHIC}

\begin{figure}
\epsfig{file=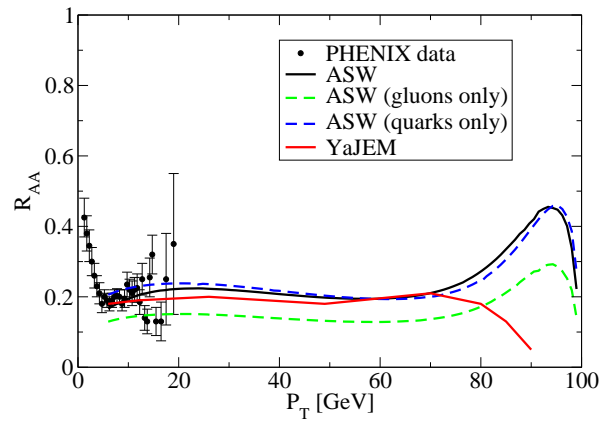, width=7.8cm}
\caption{\label{F-RAARHIC}(Color online) The nuclear suppression factor $R_{AA}$ at RHIC across the full kinematic range in 10\% central 200 AGeV Au-Au collisions. Shown are PHENIX data \cite{PHENIX_R_AA}, a calculation in the energy loss picture (ASW) with quark and gluon result shown separately, and a calculation in the medium-modified parton shower picture (YaJEM).}
\end{figure}

The nuclear suppression factor for 200 AGeV central Au-Au collisions at RHIC, calculated both in the energy loss picture (represented by the ASW model) and the medium-modified shower picture (represented by the MC code YaJEM), is shown over the full kinematic range in Fig.~\ref{F-RAARHIC} and compared with PHENIX data \cite{PHENIX_R_AA}. For the ASW calculation, the partonic $R_{AA}$ is also indicated separately for quarks and gluons.

Before discussing details of the plot, let us recapitulate the main differences between the energy loss picture as exemplified by the ASW model and the medium-modified shower picture as represented by YaJEM:

\begin{itemize}
\item ASW is derived for infinite parton energy, hence $P(\Delta E)$ is independent of the initial parton energy and the only energy dependence arises from the prescription to assign contributions where $\Delta E > E$ into an absorption term $A$. In contrast, YaJEM is a finite energy framework where the MMFF explicitly depends on the initial parton energy. In particular, within ASW there is an energy-independent transmission probability $\tilde{T}$ which bounds $R_{AA}$ from below.
\item In the energy loss picture, it is not specified what happens to the lost energy. In contrast, within YaJEM the energy lost from the leading shower partons is recovered explicitly in a low $z$ enhancement of the MMFF.
\end{itemize}

In Fig~\ref{F-RAARHIC}, these differences are apparent as follows: In the lowest $P_T$ region from 6 GeV and above, there is small rise of $R_{AA}$ with $P_T$ observed in ASW which is not seen in YaJEM. As apparent from the comparison of the ASW result for pions to the result for quarks and gluons, the rise in this region in the ASW model is driven by the transition from a gluon-dominated to a quark dominated regime --- the ASW hadronic result subsequently approaches the quark result for larger momenta. This transition is also present in YaJEM, however it is masked by the onset of the low $P_T$ enhancement, which just starts to become significant below 6 GeV and corresponds to a decreasing trend of $R_{AA}$ with increasing $P_T$. As a result, the two opposing effects roughly cancel and the YaJEM result appears  flatter than the ASW result between 6 and 25 GeV.

For higher $P_T$, there follows a region up to 50 GeV in which both the ASW and the YaJEM result descrease slightly. This can be traced back to the fact that the pQCD spectrum is not a power law, and that local power law fits result in increasing $n$ for higher $p_T$. The two curves run in parallel till $\approx 75$ GeV, then the predictions of the two models are strikingly different.

The ASW curve turns upward beyond $P_T = 75$ GeV. A comparison with Fig.~\ref{F-NPDF} shows that this has nothing to do with the final state energy loss, but reflects the Fermi motion region in the nuclear PDFs. At the kinematic boundary, the curve finally turns over to reach the transmission probability $\tilde{T}$, as all shifts in the spectrum at the kinematic boundary result in substantial suppression and the only remaining contribution are unmodified partons. In contrast, the YaJEM result shows a strong suppression from 75 GeV to the kinematic limit. This corresponds to the region $z \rightarrow 1$ in the MMFF in which suppression was always observed in Fig.~\ref{F-FFnew}, regardless of the initial energy. This suppression is strong enough to mask the enhancement from the nuclear PDF. In contrast to the ASW model, YaJEM does not include an $E$-independent transmission term, thus $R_{AA}$ becomes very small towards the kinematic limit. In this, the finite-energy nature of the suppression in YaJEM is apparent. Note that the YaJEM result cannot be computed all the way to the kinematic limit due to lack of statistics in the MC results at $z \rightarrow 1$.

It is also clear that there is no region throughout the whole kinematic range at RHIC in either model for which $R_{AA} \rightarrow 1$ could be observed. 

\section{Results for LHC}

It is then natural to expect that $R_{AA} \rightarrow 1$ could be realized by probing even higher momenta beyond the RHIC kinematic limit, e.g. by studying $R_{AA}$ at the LHC. However, in going to collisions at larger $\sqrt{s}$, not only the kinematic limit is changed, but also the production of bulk matter is increased, i.e. higher $\sqrt{s}$ corresponds to a modification of both hard probe {\em and} medium. There is however reason to expect that eventually one will find a region in which $P_T \gg T, \Delta E_{max}, \sqrt{\Delta Q^2}$ and $R_{AA} \rightarrow 1$ can be realized: The kinematic limit $\sqrt{s}/2$ increases linearly with $\sqrt{s}$. However, the medium density does not. There are different models which try to extrapolate how the rapidity density of produced matter increases with $\sqrt{s}$. The Eskola-Kajantie-Ruuskanen-Tuominen (EKRT) model is among the models with the strongest predicted increase, and has the scaling $\frac{dN}{dy} \approx \sqrt{s}^{0.574}$. Thus, the rapidity density of bulk matter increases significantly slower than the kinematic limit for increasing $\sqrt{s}$.

Although the medium lifetime may increase substantially with $\sqrt{s}$ as well, the more relevant scale is the transverse size of the medium, as high $p_T$ partons move with the speed of light and will exit the medium once they reach its edge. However, the transverse size of the medium is approximately given by the overlap of the two nuclei, and hence to 0th approximation independent of $\sqrt{s}$ (beyond, there is of course the weak logarithmic growth of all total cross sections with $\sqrt{s}$). Thus, for asymptotically high energies, the integration limits for a line integral along the parton path through the medium will not grow arbitrary large, and the integrand, i.e. the density distribution, will be the main change. All these arguments indicate that $R_{AA} \rightarrow 1$ can thus be realized for large $\sqrt{s}$ despite the increased production of bulk matter. The remaining question if the LHC energy $\sqrt{s} = 5.5$ ATeV is large enough.

\begin{figure}
\epsfig{file=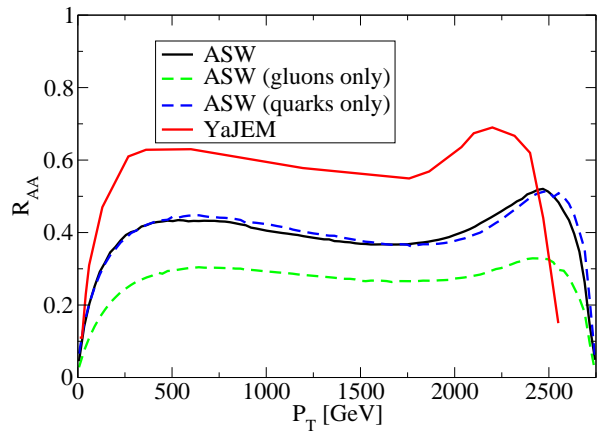, width=7.8cm}
\caption{\label{F-RAALHC}(Color online) The nuclear suppression factor $R_{AA}$ at LHC across the full kinematic range in 10\% central 5.5 ATeV Pb-Pb collisions. Shown are results in the energy loss picture (ASW) with the quark and gluon result shown separately, and a calculation in the medium-modified parton shower picture (YaJEM).}
\end{figure}

The result of the detailed calculation shown in Fig.~\ref{F-RAALHC} indicates that this is not the case. For this calculation, a hydrodynamical evolution based on an extrapolation of RHIC results using the EKRT saturation model \cite{Hydro} has been used to account for the increased medium density and lifetime. All other differences to the RHIC result are either plain kinematics, or can be traced back to the scale evolution of the MMFF.

As far as the shape of $R_{AA}(P_T)$ is concerned, the LHC predictions for both ASW and YaJEM agree, however quantitatively they differ substantially. At the heart of this difference is that ASW is an infinite energy formalism in which the larger $\sqrt{s}$ of LHC as compared to RHIC is chiefly reflected in the harder slope of the parton spectra, but not directly in $P(\Delta E)$. In contrast, within YaJEM, in addition to the harder slope of the parton spectrum, there is an explicit scale evolution of the medium effect in the MMFF (see Fig.~\ref{F-FFnew}). Since both mechanisms tend to increase $R_{AA}$, the combined effect of scale evolution and parton spectrum slope leads, all things considered, to less final state suppression in the YaJEM result.

The shape of $R_{AA}(P_T)$ can be understood by the mechanisms also observed in the RHIC case. The initial steep rise and subsequent flattening reflects the changing slope of the parton spectra. Note that the transition from gluon dominated to quark-dominated hadron production is not an issue over most of the LHC kinematical range. The final enhancement above 2 TeV is again driven by the Fermi motion region in the nuclear PDFs. Unlike in the RHIC case, at LHC kinematics the suppression obtained from YaJEM for this region is not strong enough to mask the enhancement. Finally, close to the kinematic limit, a small $R_{AA}$ is obtained.

These results indicate that there is no reason to expect that the limit $R_{AA} \rightarrow 1$ can be observed even with LHC kinematics. However, the general trend for larger $R_{AA}$ observed in the transition from RHIC to LHC indicates that the limit could be reached for asymptotically high energies over a large kinematic range, however not close to the kinematic boundary.

\section{Discussion}

So far, the nuclear suppression factor $R_{AA}$ has been observed experimentally only in a very limited kinematical region. In this region, no strong $P_T$ dependence has been observed. The main expectation of how $R_{AA}$ changes if observed over a larger kinematical range is that the suppression should eventually vanish and $R_{AA}$ approach unity. The results presented here show that this expression is too simplistic.

In particular, it is wrong to think of the shape of $R_{AA}(P_T)$ to be the result of any single cause. Instead, many effects, among them the slope change of the pQCD parton spectrum, the scale evolution of the medium modification effect, the transition from gluon-dominated to quark-dominated hadron production and also the initial state nuclear effects all influence $R_{AA}(P_T)$ in a characteristic way. Moreover, it is not sufficient to think of going to higher $P_T$ to see the lessening of the suppression --- it matters how one approaches higher $P_T$, in particular if one can push a measurement further up in $P_T$ with higher statistics, or if one measures a different system with higher $\sqrt{s}$. Based on the results presented above, it appears unlikely that the simple limit $R_{AA} \rightarrow 1$ for sufficiently high $P_T$ can be reached even at LHC kinematics.

These findings may largely be of little practical value due to the impossibility of reaching out to a substantial fraction of the kinematic limit experimentally. However, theoretically they serve well to  illustrate that even a hard probe observable like $R_{AA}$ is never 'simple' in the sense that it reflects directly tomographic properties of the medium, but rather that it is a convolution of many different effects which all need to be understood and discussed carefully. In particular, $R_{AA}$ cannot be interpreted as an observable reflecting properties of the medium causing a final state effect. The shape of the underlying parton spectrum or initial state effects are equally important to understand $R_{AA}$.
 
\begin{acknowledgments}
 
Discussions with Will Horowitz, Kari Eskola and Hannah Petersen are gratefully acknowledged. This work was supported by an Academy Research Fellowship from the Finnish Academy (Project 130472) and from Academy Project 115262. The numerical computations were carried out with generous support by Helen Caines on the {\bf bulldogk} cluster at Yale University.

\end{acknowledgments}

\end{document}